\documentclass{aa}  
\usepackage{graphicx}
\usepackage{txfonts}
\usepackage[authoryear]{natbib}
\bibpunct[; ]{(}{)}{,}{a}{}{;}
\usepackage{longtable,lscape}
\begin{document}
   \title{Stellar tracers of the Cygnus Arm\fnmsep\thanks{Partially based on observations 
collected at the 2.2-m telescope (Calar Alto, Spain),the Isaac Newton Telescope (La Palma, Spain) and 
Observatoire de Haute Provence (CNRS), France}} 
   \subtitle{II. A young open cluster in \object{Cam OB3}}

   \author{I. Negueruela
          \inst{1}
          \and
          A. Marco
          \inst{1}
          }

   \offprints{I. Negueruela}

   \institute{Departamento de F\'{i}sica, Ingenier\'{i}a de Sistemas y
     Teor\'{i}a de la Se\~{n}al. Escuela Polit\'ecnica
     Superior. Universidad of Alicante. Apdo.99
     E-03080. Alicante. (Spain)\\ 
              \email{ignacio@dfists.ua.es}}

   \date{}

% \abstract{}{}{}{}{} 
% 5 {} token are mandatory
 
\abstract{Cam OB3 is the only defined OB association believed to belong to
  the Outer Galactic Arm or Cygnus Arm. Very few members have been
  observed and the distance 
  modulus to the association is not well known.} 
% aims heading ()
{We attempt a more complete description of the population of Cam OB3
  and a better determination of its distance modulus.} 
{We present $uvby$ photometry of the area surrounding the O-type
stars \object{BD $+56\degr$864} and LS~I~$+57\degr$138, finding a
clear sequence of early-type stars that define an uncatalogued open
cluster, which we call Alicante~1. We also present spectroscopy of
stars in this cluster and the 
surrounding association.}
{From the spectral types for 18 very likely members of the association
  and $UBV$ photometry found in the literature, we
  derive individual reddenings, finding a extinction law close to
  standard and an average distance modulus $DM=13.0\pm0.4$. This value is in
  excellent agreement with the distance modulus to the new cluster
  Alicante~1 found by fitting the photometric sequence to the ZAMS. In
spite of the presence of several O-type stars, Alicante~1 is a very
sparsely populated open cluster, with an almost total absence of early
B-type stars.}
{Our results definitely confirm Cam~OB3 to be located on the Cygnus
  Arm and identify the first open cluster known to belong to the association.}

\titlerunning{New open cluster in Cam OB3}

\maketitle 
%
%________________________________________________________________

\section{Introduction}

The structure of the Milky Way is still the subject of
controversy. While many authors have argued for  a four-armed Galaxy
\citep[e.g.,][]{val05,rus03}, two-armed models have also remained
popular \citep[e.g.,][]{fer01}. Recent results based on star counts
from the GLIMPSE survey \citep{ben05}
favour a two-armed Galaxy, with the Scutum-Crux (i.e., the second arm
towards the Galactic Centre from the position of 
the Sun or$-$II) and the Perseus (the first arm towards the outside
from the position of the Sun or $+$I) arms
as main features. This is a surprising result, as previous authors
had considered the Sagittarius-Carina 
($-$I) Arm and the Norma ($-$III or internal) Arm as the
main features \citep[e.g.,][]{gag}.

Unfortunately, our view from the Galactic periphery is not the most
adequate to study Galactic structure. For example, the Perseus Arm,
now believed to be a major spiral feature, cannot be traced beyond
Galactic longitude $l\approx140\degr$, though some tracers seem to be
found around $l\approx180\degr$. This could reflect a flocculent
morphology for the outer Milky Way \citep[e.g.,][]{qui02}. 

Beyond the Perseus Arm, molecular clouds definitely delineating an
outer arm, generally referred to as the Cygnus Arm, are readily
visible in CO surveys all over the first 
Galactic quadrant \citep{dam01}. In
most models,  the Cygnus Arm is assumed to be the 
continuation of the Norma Arm.
In a previous paper \citep[][ henceforth Paper I]{uno}, we presented
spectroscopy for several stars that could belong to the Cygnus
Arm. Tracers were particularly reliable in the region
$l\approx140\degr$\,--\,$180\degr$, where the Perseus Arm is not present
along the line of sight. Several stars around $l=147\degr$, were
assumed to belong to the association \object{Cam OB3}.

\defcitealias{uno}{Paper~I}

The existence of \object{Cam OB3} has sometimes been considered
doubtful, as the density of members is not very high and no previously
catalogued open clusters are contained. Using data in the literature for
6 likely members, \citet{humphreys} centred it at ($l=147\degr$,
$b=+3.0$) and derived $DM=12.6$. \citet{haug} obtained $UBV$
photometry of a larger number of OB stars in the first volume of the
Luminous Star catalogue and, based on estimated distances, considered
the existence of  \object{Cam OB3} certain. In \citetalias{uno}, we
obtained spectroscopy for 10 likely members and found that their
calculated distance moduli concentrated very tightly around
$DM=13.0$. This $DM$ is one and a half magnitude
higher than those to \object{Per OB1} and \object{Cas OB6}, the
tracers of the Perseus Arm closest in the sky, implying that
\object{Cam OB3} is clearly too far away to be on the Perseus Arm.

During this work, we noticed that sky images of the area between the 
O-type stars \object{BD $+56\degr$864} (O6\,Vnn, double-lined
spectroscopic binary) and \object{LS I $+57\degr$138} (O7\,V) showed a
very high concentration of moderately bright stars. In this paper, we present
photometry of the area, clearly showing the presence of an early-type
star sequence identifiable as a small open cluster. We also present a
rather larger spectroscopic sample of possible members of Cam~OB3,
allowing us a much better definition of the association.

\section{Observations and data}

\subsection{Photometry}
We obtained Str\"omgren $uvby$ photometry of the region around the
star \object{BD~$+56\degr$864} using BUSCA attached to the 2.2-m
telescope at the 
Calar Alto Observatory (Almeria, Spain) on the nights of 24\,--\,26 October
2002.  

The instrument splits the light into 4 channels, each equipped with a
CCD camera, which can take images at the
same time in different filters. We used the 3 bluer channels to obtain
images with the $uvby$ filters. Each camera covers a field of view
of $12\arcmin \times 12\arcmin$ and has a pixel scale of $0\farcs176$.
Images from the area  were taken using 3 series of different exposure
times to obtain accurate photometry for a magnitude range. These
exposure times are given in Table~\ref{table:1}.  

%-----------------------------------------------------------------------

\begin{table}
\caption{Log of the photometric observations taken at the 2.2-m on
  October 2002. All times are in seconds.}
\label{table:1}
\centering
\begin{tabular}{c c c}
\hline\hline
Alicante 1& ${\rm RA} = 03\,{\rm h}\: 59\,{\rm m}\: 18.29\,{\rm s}$ & DEC = $+57\,^{\circ}\, 14\arcmin\, 13.7\arcsec$ \\
&(J2000)&(J2000)\\
\end{tabular}
\begin{tabular}{c c c c}
\hline\hline
Filter & Long times & Intermediate times & Short times \\
\hline
$u$ & 1400 & 100 & 5 \\
$v$ & 400 & 40 & 2 \\
$b$& 400 & 40 & 2 \\
$y$ & 400 & 40 & 2 \\
\hline
\end{tabular}
\end{table}

%-----------------------------------------------------------------------

The reduction of the images was done with IRAF\footnote{IRAF is
  distributed by the National Optical Astronomy Observatories, which
  are operated by the Association of Universities for Research in
  Astronomy, Inc., under cooperative agreement with the National
  Science Foundation} routines for the bias and flat-field
corrections. The photometry was obtained by point-spread function
(PSF) fitting using the DAOPHOT package \citep{stetson1987} provided
by IRAF. The apertures used were the same for standard and target
stars. In order to construct the PSF empirically, we manually selected
bright stars (typically 25 stars) over the whole field as good candidates
to be PSF stars. Once we had the list of candidate PSF stars, we determined an
initial PSF by fitting the best function among the 5 options available
in the PSF
routine inside DAOPHOT. We chose the PSF to be variable in order 2
across the frame to take into account the systematic pattern of PSF
variability with position on the chip.

We selected secondary standard stars from \citet{mar01}.
We performed atmospheric extinction corrections and solved the
transformation equations following the procedure described in
\citet{mar01}. The precision of the photometry is calculated as the
dispersion of the mean of the difference (transformed value $-$ catalogue
value). The values for this run are given in Table~\ref{table:2}.

%----------------------------------------------------------------------- 

\begin{table}
\caption{Mean catalogue minus transformed values ($D$) for the standard
  stars and their standard deviations in $V$, $(b-y)$ and $c_{1}$}
\label{table:2}
\centering
\begin{tabular}{l c c c}
\hline\hline
&$V$&$(b-y)$&$c_{1}$\\
\hline
$D$&0.00&0.00&0.00\\
$\sigma$&0.02&0.03&0.03\\
\hline
\end{tabular}
\end{table}   
%----------------------------------------------------------------------- 

\begin{figure}
\resizebox{\columnwidth}{!}{\includegraphics{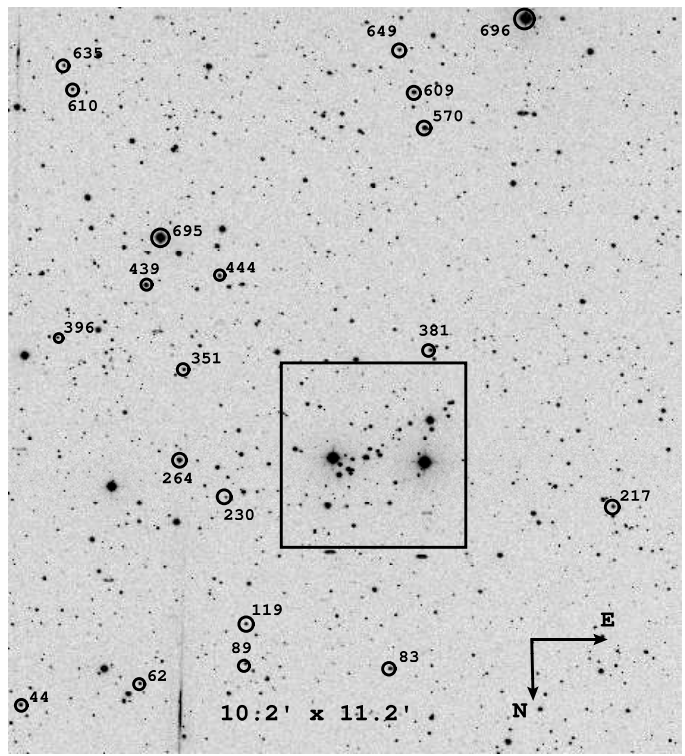}}
\centering
\caption{Finding chart for stars belonging to the blue sequence in
  Fig.~\ref{fig:rawhr} (possible cluster members). $XY$ positions are
  listed in Table~\ref{tab:rawphot}, where they are correlated to RA and
  DEC. The origin of coordinates is located at the bottom left corner
  of the image. The area inside the square is displayed in the following
  figure. North is down and East is right. This is a $y$-band
    image taken with BUSCA. The field of view is $10\farcm2 \times
    11\farcm2$ and the central coordinates  are ${\rm RA} = 03\,{\rm
      h}\: 59\,{\rm m}\: 19\,{\rm s}$ \& DEC = $+57\,^{\circ}\,
    11\arcmin\, 53\arcsec$ in epoch J2000.}  
\label{fig:full} 
\end{figure}

\begin{figure}
\resizebox{\columnwidth}{!}{\includegraphics{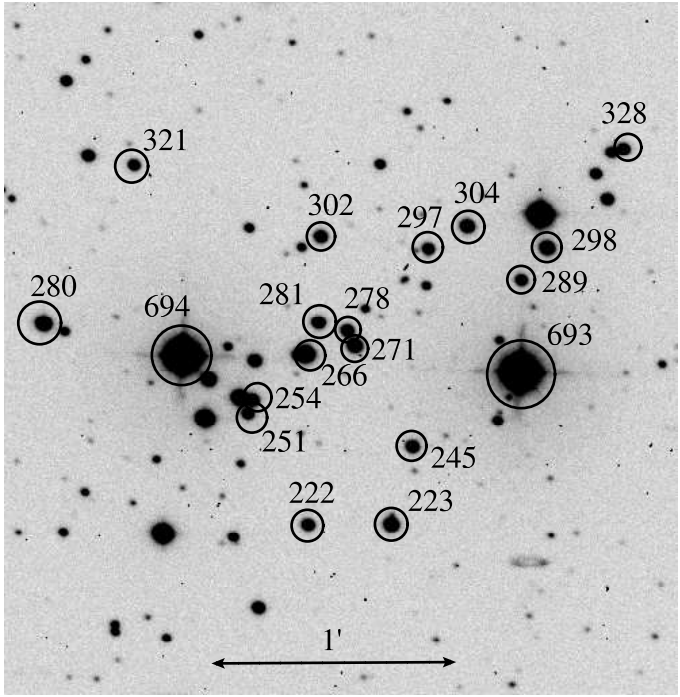}}
\centering
\caption{Finding chart for likely cluster members in the central
  concentration. $XY$
  positions are listed in Table~\ref{tab:rawphot}, where they are
  correlated to RA and DEC. The origin of coordinates is located at
  the bottom left corner of Fig.~\ref{fig:full}.} 
\label{fig:zoom}
\end{figure}

In Fig.~\ref{fig:full}, we show the area observed. Likely members of
the cluster are numbered. As this is a newly discovered
cluster, we use our own numbering system. In Fig.~\ref{fig:zoom}, we
zoom in on the area of the image 
marked with a square, which contains the majority of cluster
members. In Table~\ref{tab:rawphot}, we give $X$ and $Y$ positions for
stars in Fig.~\ref{fig:full} and~\ref{fig:zoom}, together with their identification in the
2MASS catalogue an their coordinates (right ascension, RA, and
declination, DEC) in epoch J2000. 
In Table~\ref{table:4}, we give the values of $V$, $(b-y)$ and $c_{1}$ for
these stars.

%-----------------------------------------------------------------------

%-----------------------------------------------------------------------

%-----------------------------------------------------------------------

\subsection{Spectroscopy}

\begin{figure}
\centering
\resizebox{\columnwidth}{!}{\includegraphics[angle =-90]{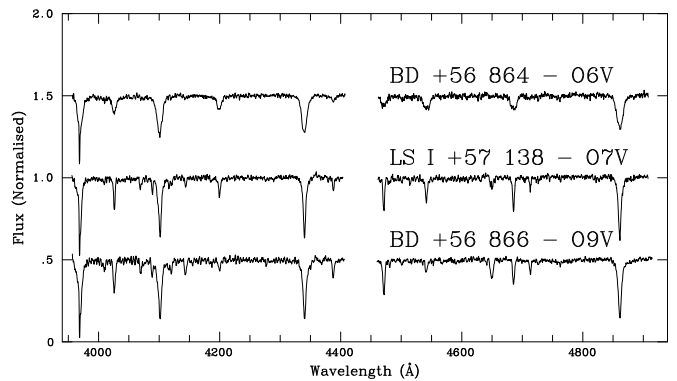}}
  \caption{OHP 1.52-m spectra of the three O-type stars in the area
    that we investigated. The identifications in
    Fig.~\ref{fig:full}~and~\ref{fig:zoom} are
    BD~$+56\degr$864 = \#693, LS~I~$+57\degr$137 = \#695 and
    BD~$+56\degr$866 = \#696. The
   small gap around $\lambda = 4420$\AA\ indicates the division
   between the two poses. Details of the observations are given in
   \citetalias{uno}.} 
   \label{fig:camob3}
\end{figure}

Observations of several stars in \object{Cam OB3}, including the three
O-type stars likely connected with the new cluster, were presented in
\citetalias{uno}. These spectra had been obtained with the {\em Aur\'{e}lie}
spectrograph on the 1.52-m telescope at the Observatoire de Haute
Provence (OHP) during  runs on 18th--22nd January 2002 and 25th--28th
February 2002 (see \citetalias{uno} for details). The spectra of the
  three O-type stars in the cluster 
area are shown in Fig.~\ref{fig:camob3}. 

A few objects, which had not been observed over the whole
3960\,--\,4910\AA\ range, were not
included in \citetalias{uno}, as their spectral types were not
certain, in most cases because the spectra clearly showed lines
corresponding to two stars. These observations are listed in
Table~\ref{tab:ohp}. 

%-----------------------------------------------------------------------

\begin{table}[ht]
\caption{Observations from the OHP 1.52-m telescope that were not
  included in Paper~I.} 
  \begin{center}
\begin{tabular}{lccc}
Name & Date& Exposure & Wavelength\\
& &Time (s) &  Range \\
\hline
HD 237204 &  28/02/02 & 900 & $3960-4410$\AA\\
BD $+55\degr$837 & 25/02/02 & 1500 & $4460-4910$\AA\\
BD $+55\degr$838 & 25/02/02  & 1500 & $4460-4910$\AA\\
LS I $+57\degr$137 & 21/01/02 & 1800 &$4460-4910$\AA\\
LS V $+56\degr$59 & 20/01/02 & 1800 &$3960-4410$\AA\\
\end{tabular}
\end{center}
 \label{tab:ohp}
\end{table}

%-----------------------------------------------------------------------

Further spectra of these and other stars in \object{Cam OB3} were
obtained with the Intermediate Dispersion Spectrograph (IDS) at the
2.5-m Isaac Newton Telescope (INT) in La Palma (Spain) during a run on
23th--26th July 2002. The instrument was equipped with the 235-mm
camera, the R1200Y grating and the Tek\#5 CCD. This configuration
gives a dispersion of $\sim0.8$\AA/pixel over the $4050 -
4900$\AA\ range.  The complete log of these observations is given
in Table~\ref{tab:intjul}. 

%-----------------------------------------------------------------------

\begin{table}[ht]
\caption{Possible members of \object{Cam OB3} observed from the INT in
  July 2002.} 
  \begin{center}
\begin{tabular}{lccc}
LS & Other & Date& Exposure \\
Number& name & &Time (s) \\
\hline
I $+56\degr$90 & $-$          &26/07/02 &800\\
I $+56\degr$92 & $-$          &23/07/02 &450\\
I $+56\degr$94 & $-$          &25/07/02 &500\\
I $+56\degr$98 & HD~237204       &26/07/02 &300\\
I $+55\degr$53 & $-$          &26/07/02 &800\\
I $+55\degr$54 & $-$          &25/07/02 &500\\
I $+55\degr$55 &BD\,$+55\degr$837 &23/07/02 &300\\
I $+55\degr$57 & $-$          &26/07/02 &900\\
I $+55\degr$58 &BD\,$+55\degr$838 &23/07/02 &300\\
I $+55\degr$47 & $-$          &25/07/02 &900\\
V $+56\degr$59 & $-$          &25/07/02 &400\\
I $+57\degr$136 & $-$         &25/07/02 &600\\
I $+57\degr$137 & $-$         &26/07/02 &720\\
I $+57\degr$140 & $-$         &26/07/02 &500\\
\end{tabular}
\end{center}
 \label{tab:intjul}
\end{table}
%-----------------------------------------------------------------------

Finally, spectra of possible members of the new cluster have been
obtained during
different runs. Some were observed with the INT on 2002 October
25\,--\,28. The instrument was 
equipped with the 235-mm
camera and the EEV \#10 CCD. We used the R1200B grating for
the brightest members (covering 3900\,--\,5200\AA) and the R400V grating
for faint stars (covering 3500\,--\,7200\AA). Other members were
observed on Aug 29th 2008 with the 2.6-m Nordic Optical Telescope (La
Palma) and ALFOSC. We used grism \#14 to cover the 3900\,--\,6800\AA\ 
range. Some of these spectra are shown in
Figures~\ref{fig:bins}~and~\ref{fig:al1members}. 

All the spectroscopic data have been reduced with the {\em Starlink}
packages {\sc ccdpack} \citep{draper} and {\sc figaro}
\citep{shortridge} and analysed using {\sc figaro} and {\sc dipso}
\citep{howarth}. 

%-----------------------------------------------------------------------

%-----------------------------------------------------------------------

\section{Results}

\subsection{Photometry}

\subsubsection{Observational HR diagram}

\begin{figure}[ht]
   \resizebox{\columnwidth}{!}
   {\includegraphics{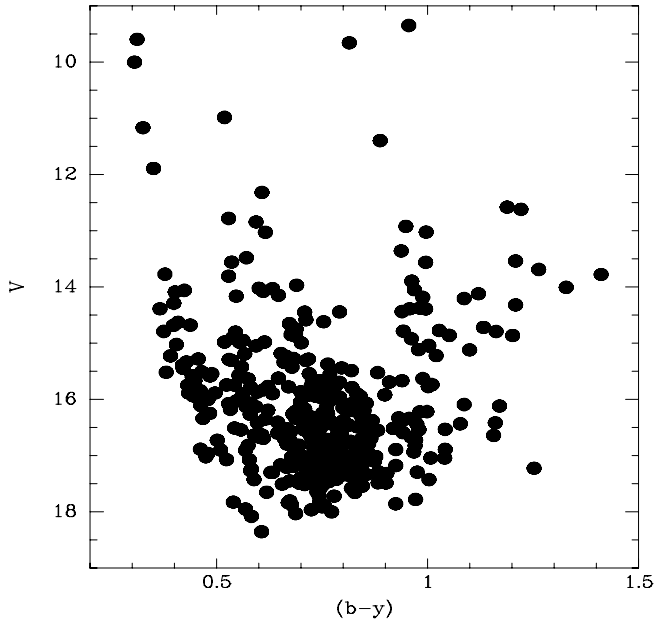}}
   \caption{$V/(b-y)$ diagram for all 
            stars in the field. We identify the bluer sequence as the
            new cluster Alicante~1.
           }
   \label{fig:rawhr}
    \end{figure}

\begin{figure}[ht]
   \resizebox{\columnwidth}{!}
   {\includegraphics{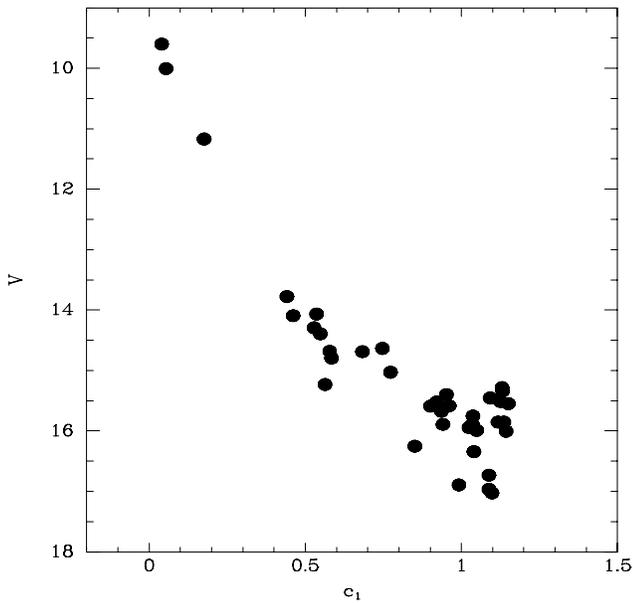}}
   \caption{$V/c_{1}$ diagram for the stars 
            stars  identified as
            possible cluster members in Fig.~\ref{fig:rawhr}.
           }
   \label{fig:rawc1}
    \end{figure}

We start by plotting the $V/(b-y)$ diagram for all stars in the field (see
Fig.~\ref{fig:rawhr}). The diagram shows evidence for different
stellar populations, with at least two well-defined sequences. 
To the left of the diagram, there is a clear
sequence with bluer colours, $(b-y)\la0.5$. Catalogued OB stars lie at
the top of this sequence  and hence it seems sensible to assume
  that this sequence corresponds to an associated population. To check
this hypothesis, we select all the stars belonging to this
sequence and plot their $V-c_{1}$ diagram
(see Fig.~\ref{fig:rawc1}). The stars also form a clear sequence in
this diagram, with colours typical of reddened
early type stars. Stars belonging to the sequence are marked in
Figures~\ref{fig:full} and~\ref{fig:zoom}. It is obvious that the
vast majority of the stars in the sequence are grouped together 
in a small area surrounding the two mid O-type stars. In view of the spatial
concentration of the photometric sequence, we conclude that these
stars are members of a previously uncatalogued open cluster in this
region, which we will provisionally call Alicante~1.

%-----------------------------------------------------------------------

%-----------------------------------------------------------------------

%-----------------------------------------------------------------------

The procedure used may leave out some cluster members, but guarantees
  that most candidates selected are cluster members. The sequence
  shows little spread in $E(b-y)$ (see following section) and therefore
  it is unlikely that main-sequence members with higher reddenings are
  left out. Moreover, \citep{zda05} argue that most of the reddening
  in this direction originates close to the Sun. The small age of
  Alicante~1 (see Section~\ref{sec:cluster}) precludes the existence
  of evolved members. Pre-main-sequence (PMS) members are likely to
  exist, and our selection procedure is likely to leave out most of
  them. Finding out these PMS objects from photometry alone is very
  difficult, as they will not have typical colours and the cluster is
  sparse and mixed with a large foreground and background
  population. A search for 
  emission-line objects may produce a sample of this population.
 
\subsubsection{Determination of the distance}
\label{sec:dist}

Our selection of candidates leaves 38 stars as possible members of the
new cluster Alicante~1. Now we can use the photometric analysis to
determine its distance. First, we need to deredden the values for $V$,
$(b-y)$ and $c_{1}$. For this purpose, we calculate individual 
reddenings following the procedure described by \citet{craw70} for
B-type stars: we use the observed $c_{1}$ to 
predict the first approximation to 
$(b-y)_{0}$ with the expression $(b-y)_{0}=-0.116+0.097c_{1}$. Then we
calculate $E(b-y)=(b-y)-(b-y)_{0}$ and use $E(c_{1})=0.2E(b-y)$ to 
correct $c_{1}$ for reddening $c_{0}=c_{1}-E(c_{1})$. The intrinsic
color $(b-y)_{0}$ is now calculated by replacing $c_{1}$ with $c_{0}$ in
the above equation for $(b-y)_{0}$. Three iterations are enough to
reach convergence in the process. The final values of $E(b-y)$,
$E(c_{1})$ and $c_{0}$ are given in Table~\ref{table:4}. We
note that this procedure assumes that the reddening law is the average
for nearby stars. We will see in Section~\ref{sec:extlaw} that this
assumption is justified.

The reddening is quite constant across the cluster. The average value
is $E(b-y)=0.48$ for 38 possible members, with a dispersion of only
$\sigma=0.03$, comparable to the precision of the $(b-y)$ colour in
our photometry. No star deviates by more than $2\sigma$ from the
average value, which corresponds to $E(B-V)=0.67$.

\begin{figure*}[ht]
   \centering
   \includegraphics[width=\columnwidth]{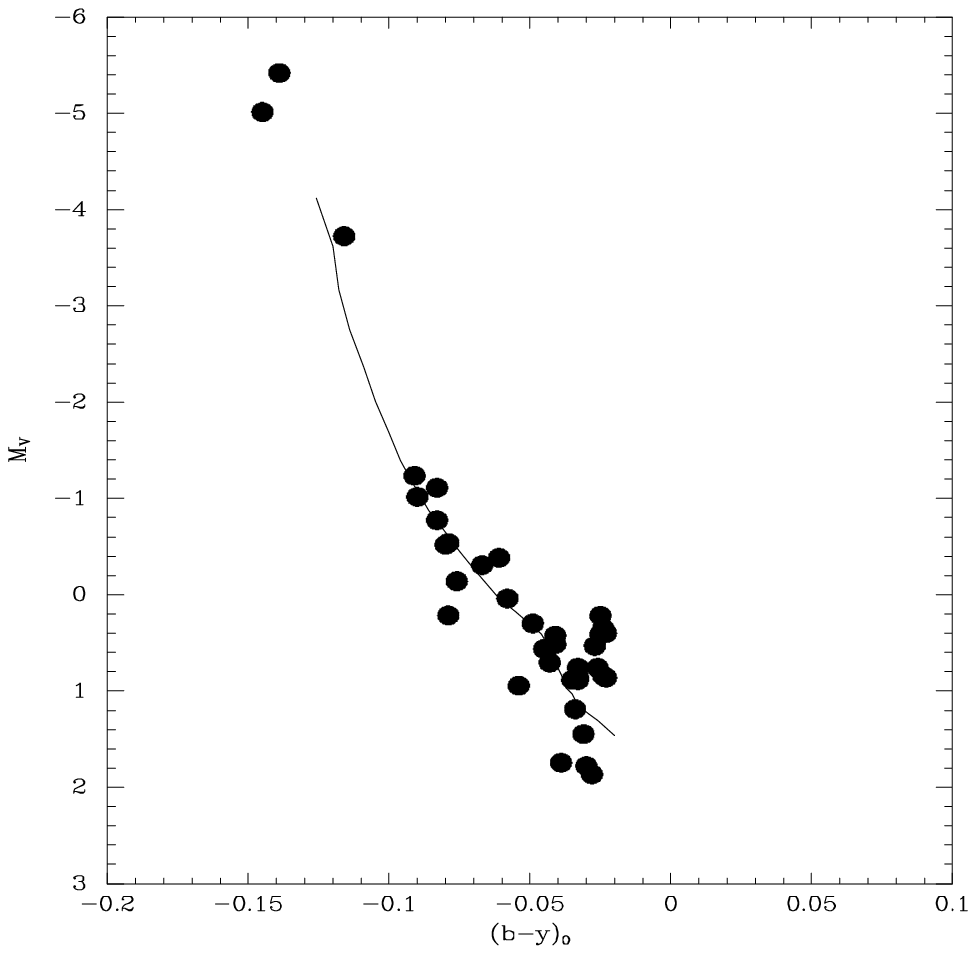}
   \includegraphics[width=\columnwidth]{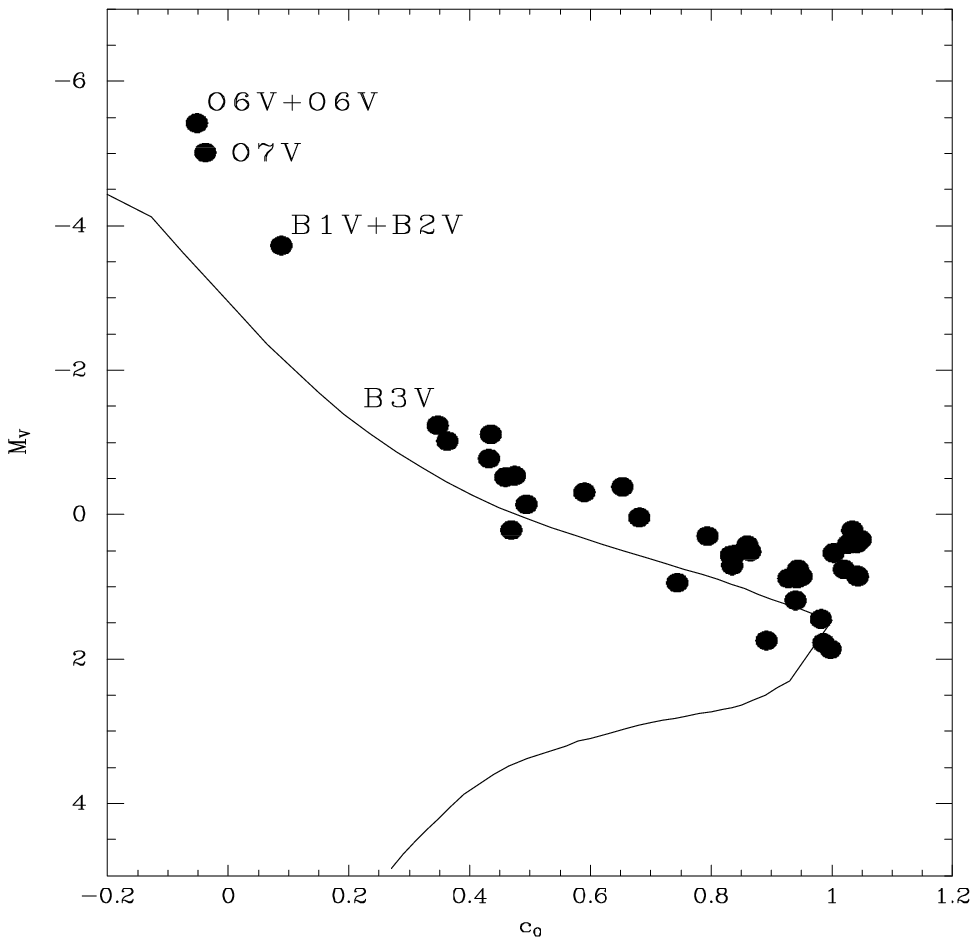}
   \caption{{\bf Left: }Dereddened $M_{V}/(b-y)_{0}$ diagram for
     cluster members. {\bf Right: } Dereddened $M_{V}/c_{0}$ diagram
     for members. The spectral types for the brightest members are
     indicated. In both panels, the line represents the ZAMS from
     \citet{perry1987}.} 
   \label{fig:real}
    \end{figure*}

Next, we plot the $V_{0}/(b-y)_{0}$ and $V_{0}/c_{0}$ diagrams (see
Fig.~\ref{fig:real}) and we calculate the distance modulus $DM$ that
gives the  best fit to the ZAMS \citep{perry1987} in both
diagrams. The best fit for the distance modulus is 
$DM=13.0\pm0.2$, using the sequence of late B and
early-A stars. In the $V_{0}/c_{0}$ diagram, it is clear that the
three brightest members are displaced vertically from their expected
positions. This is a suggestion of a binary nature, and we will see in
Section~\ref{sec:members} that two of the three are spectroscopic
binaries. Their position in the diagram is compatible with the same
$DM$ if this is taken into account. In Table~\ref{table:4}, we list
the dereddened photometric indices and the derived values for absolute
magnitudes calculated as $M_{V}=V_{0}-13.0$. 

%-----------------------------------------------------------------------

%-----------------------------------------------------------------------

\subsection{Spectroscopy}

\subsubsection{Cluster members}
\label{sec:members}

\begin{figure}
\centering
\resizebox{\columnwidth}{!}{\includegraphics[angle =-90]{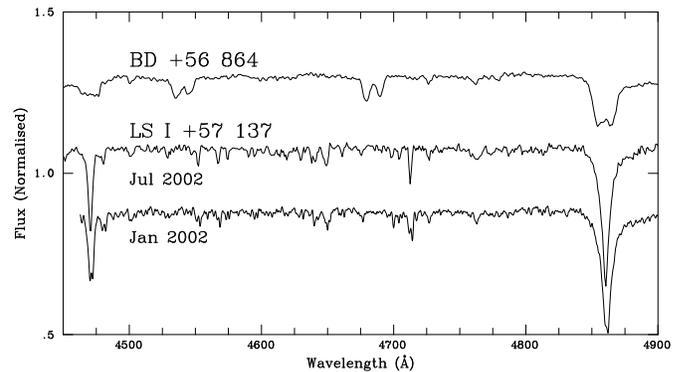}}
  \caption{Spectra of two of the brightest members of Alicante~1
    showing that they are double-lined spectroscopic
    binaries. BD~$+56\degr$864, originally classified as O6\,Vnn, has
    two similar components. In the case of LS~I~$+57\degr$137, the
    spectrum from July 2002 shows a single component of spectral type
    B1.5\,V. However, the January 2002 spectrum clearly shows double
    \ion{He}{i} and \ion{Mg}{i}~4481\AA\ lines and weak second
    components in other metallic lines.}
   \label{fig:bins}
\end{figure}

The brightest member of Alicante~1 is BD~$+56\degr$864
(\#693). Several spectra of this object show double lines (see
Fig.~\ref{fig:bins}). The two 
components have similar spectral types, around O6\,V. This system is a
short period binary and will be studied in detail in forthcoming
work. Double lines are also evident in the OHP spectrum of \object{LS
  I $+57\degr$137} (\#695). The secondary shows weaker metallic lines and
therefore appears to be slightly later than the primary. The INT
spectrum can be classified as B1.5\,V, but, given the double lines,
this may be a combination of a $\sim$B1\,V and a $\sim$B2\,V star.

The expected absolute magnitude of an O6\,V star is $M_{V}=-4.9$
\citep{martins} and
so the observed $M_{V}=-5.4$ of BD~$+56\degr$864 is compatible with the presence of a secondary almost as bright as the primary. The expected magnitude of a
B1\,V star is $M_{V}=-3.2$ and again the observed $M_{V}=-3.7$ of
LSI~$+57\degr$137 is 
compatible with the presence of a bright secondary. The spectrum of
LS~I~$+57\degr$138 (\#694) (O7\,V), the second brightest member, does
not show any evidence for a secondary component (see
Fig.~\ref{fig:camob3}). The observed $M_{V}=-5.0$ is also a
bit too high for a single star, but within the expected dispersion of
the calibration ($M_{V}=-4.6$ according to \citealt{martins}).

To the South of the field, lies BD~$+56\degr$866. Though this star is
marked in Fig.~\ref{fig:full} as \#696, it is too close to the edge of
the chip, and we could not obtain its photometry. However, its
published photometry clearly shows it to be a member of Cam~OB3 (see
Table~\ref{tab:data}) and so it is almost certainly a cluster
member. The $V/(b-y)$ diagram (Fig.~\ref{fig:rawhr}) shows a star with
$V\approx12$ that seems to belong to the cluster sequence, but is not
present in Fig~\ref{fig:rawc1}. This object was very close to the edge
of the chip in the short $u$ exposure and we have no $c_{1}$ index for
it. It is identified as GSC 03725-00486, and its 2MASS colours show it
to be an early-type star, though its $(J-K_{{\rm S}})$ is smaller than
those of cluster members. We took a spectrum of this object, which
turns out to be an A giant and so not a cluster member.

\begin{figure}
\centering
\resizebox{\columnwidth}{!}{\includegraphics[angle =-90]{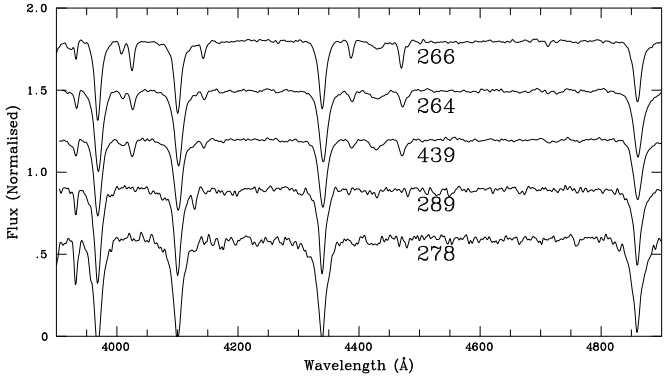}}
  \caption{Spectra of some cluster members. Stars \#264, \#266 and
\#439 are the brightest members in the photometric sequence after the
catalogued OB stars, Stars \#289 and \#278 are fainter members
observed by chance. See text for the spectral classifications.}
   \label{fig:al1members}
\end{figure}

Figure~\ref{fig:al1members} shows the spectra of other likely
members. The brightest non-catalogued member,\#266, has spectral type
B3\,V.  The next brightest stars, \#264  and
\#439, have spectral types B4\,--\,B5\,V,
in very good agreement with their positions in the photometric
diagrams. Other stars observed by chance are \#289 and \#278. Star
\#289 is located in the cluster core. It
has weak \ion{Ca}{ii}~3933\AA, 
compatible with the purely interstellar line seen in the B-type
stars. The prominent \ion{Si}{ii}~4128\AA\ doublet then suggest that
this is a chemically peculiar Bp (Si) star. The spectral type is
$\sim$B9, and the object is likely to be a Bp member. Star \#278 is
also in the cluster core. It has a spectral type around
A1\,III--IV. As its reddening is typical of cluster members and
  its position in 
  the $M_{V}/c_{0}$ diagram is compatible with the same distance
  modulus as the rest of the cluster, it may only be a chance
  alignment of a nearby unrelated star or a PMS cluster member. PMS
  stars without emission lines have been identified in
  clusters with low background contamination, such as NGC~1893
  \citep{neg07}. As a matter of fact, \#278 is part of a small clump
  of objects with $c_{0}\approx1.0$, lying slightly above the rest of
  the sequence (Fig.~\ref{fig:real}), and this discussion may apply to
  all of them.

%-----------------------------------------------------------------------

%-----------------------------------------------------------------------

%-----------------------------------------------------------------------

\subsubsection{Other members of Cam~OB3}

For all the candidate members of Cam~OB3, we derive spectral types
following the prescriptions described in \citetalias{uno}. The
resulting spectral types are listed in Table~\ref{tab:data}. We will use this
information to derive the reddening and distance to the stars in the
next Section. 
In the OHP spectrum, \object{LS V $+56\degr$59} clearly appears as a
double-lined spectroscopic binary.  The \ion{H}{i} and \ion{He}{i}
lines of the secondary spectrum are weaker than those of the primary,
while the metallic lines appear similar in strength.
HD~237204 is
also likely to be a double-lined spectroscopic binary, as all metallic
lines appear very broad in the two spectra available. \citet{rub65},
however, did not detect any variability in its radial velocity. 
The lines of \object{LS I $+57\degr$140} are also broad and
asymmetric, but the presence of a secondary spectrum is not certain.
%\object{LS I $+55\degr$57} may be a bit C-defficient

\subsection{Reddening and distance to Cam~OB3}
\label{sec:extlaw}

In order to determine the reddening and distance to the isolated stars, 
we have used $UBV$ photometry from the literature. As in \citetalias{uno}, 
we have taken \citet{haug} as our prime source of 
photometry, supplemented by \citet{hil56}. Both authors use the same photometric
system. In addition, we have collected $JHK_{\rm S}$ photometry from the 2MASS 
catalogue \citep{skru06}. The photometry has been used as input for the 
$\chi^{2}$ code for parametrised modelling and characterisation of
photometry and spectroscopy {\sc chorizos}
implemented by \citet{maiz04}. This code generates synthetic
photometry from a stellar model and convolves it with extinction laws from
\citet{cardelli}. A fit to the real photometric data determines the values
of $R$ and $E(B-V)$ that would most likely result in the observed
photometry for a star with the given spectrum. In
Table~\ref{tab:data}, we list the spectral types derived for all
candidate members of Cam~OB3, the photometry collected from the
literature and the values obtained for $R$ and $E(B-V)$. We consider
all the candidate members from \citetalias{uno} and the objects listed in
Tab.~\ref{tab:intjul}.

\begin{table*}[ht]
\caption{Candidate members of \object{Cam OB3}, with their spectral
  types, photometry from the literature and derived parameters.}
  \begin{center}
\begin{tabular}{lcccccccccc}
LS & $l$& $b$&Spectral & $V$ &  Ref.$^{(1)}$ & $R$& $E(B-V)$ &$m_{V}$ &$DM$\\
Number&&&  Type &  & & &\\
\hline
I $+55\degr$47 &$143\fdg5$&$-1\fdg5$&O9\,III&11.25 &h &      	 $2.88\pm0.08$  &  $1.35\pm0.03$  & $7.34\pm0.04$ &  12.6\\
I $+56\degr$90 &$145\fdg9$&$+1\fdg9$&B1.5\,V(+)&12.24 &h &   	 $2.74\pm0.12$  &  $0.81\pm0.03$  & $9.98\pm0.04$ &  12.8\\
I $+57\degr$136&$146\fdg0$&$+2\fdg9$&O8.5\,V&10.88&h&        	 $2.92\pm0.15$  &  $0.66\pm0.03$  & $8.90\pm0.04$ &  13.1\\
I $+56\degr$92 &$146\fdg0$&$+2\fdg4$&B5\,IIe&10.26 &h&	         $3.38\pm0.14$  &  $0.86\pm0.03$  & $7.31\pm0.04$ &  12.7 \\
I $+57\degr$140&$146\fdg1$&$+3\fdg5$&B0.2\,IV(+) &11.02 &h&  	 $2.91\pm0.17$  &  $0.60\pm0.03$  & $9.22\pm0.04$ &  13.7\\
I $+56\degr$94 &$146\fdg2$&$+2\fdg8$&B1.5\,IV& 11.77&h&      	 $3.00\pm0.25$  &  $0.46\pm0.03$  & $10.37\pm0.04$&  13.7\\ 
I $+57\degr$137&$146\fdg3$&$+3\fdg1$&B1.5\,V+&11.23 &h&      	 $2.90\pm0.20$  &  $0.52\pm0.03$  & $9.66\pm0.04$ &  12.5\\
I $+57\degr$138&$146\fdg3$&$+3\fdg1$&O7\,V &10.09&h&         	 $3.20\pm0.19$  &  $0.56\pm0.02$  & $8.27\pm0.04$ &  12.9\\
I $+57\degr$139&$146\fdg3$&$+3\fdg1$&O6\,V+ &9.67&h&         	 $2.74\pm0.18$  &  $0.56\pm0.02$  & $8.14\pm0.04$ &  13.1\\
I $+56\degr$97 &$146\fdg4$&$+3\fdg1$&O9\,V&10.33 & h &      	 $2.94\pm0.17$  &  $0.64\pm0.03$  & $8.42\pm0.04$ &  12.5\\
I $+56\degr$98 &$146\fdg6$&$+3\fdg0$&B0.5\,V+ &9.17&h&       	 $2.86\pm0.19$  &  $0.54\pm0.03$  & $7.59\pm0.04$ &  11.1 \\
I $+55\degr$54 &$146\fdg9$&$+1\fdg8$&B0.2\,IV&11.77 &h &     	 $2.92\pm0.12$  &  $0.85\pm0.03$  & $9.26\pm0.04$ &  13.8\\
I $+55\degr$53 &$147\fdg7$&$+1\fdg5$&B3\,III$^{*}$&11.83 &h& 	 $3.11\pm0.11$  &  $1.00\pm0.03$  & $8.69\pm0.04$ &  11.7\\
I $+55\degr$55 &$147\fdg0$&$+2\fdg0$&BC1.5\,Ib&9.59&h&	         $2.97\pm0.13$  &  $0.88\pm0.03$  & $6.94\pm0.04$ &  12.7\\
I $+56\degr$99 &$147\fdg1$&$+3\fdg0$&O9.5\,Iab& 9.00 &h&     	 $3.13\pm0.14$  &  $0.74\pm0.02$  & $6.64\pm0.04$ &  12.9\\
V $+56\degr$56 &$147\fdg4$&$+3\fdg9$&B5\,Ia&7.99& hi &       	 $3.51\pm0.19$  &  $0.64\pm0.03$  & $5.73\pm0.04$ &  12.9\\
I $+55\degr$57 &$147\fdg5$&$+1\fdg8$&B1\,IV$^{*}$&12.19&h&  	 $2.96\pm0.14$  &  $0.78\pm0.03$  & $9.81\pm0.04$ &  13.6\\
I $+55\degr$58 &$147\fdg6$&$+1\fdg9$&B2.5\,Iab &9.29 &h &    	 $2.98\pm0.11$  &  $1.00\pm0.03$  & $6.28\pm0.04$ &  12.8\\
V $+55\degr$11 &$147\fdg6$&$+2\fdg7$&B6\,Ia &8.72 &hi &      	 $3.15\pm0.14$  &  $0.84\pm0.03$  & $6.04\pm0.04$ &  13.0\\
V $+56\degr$59 &$149\fdg7$ &$+5\fdg3$&B1\,V+&$10.92^{*}$&$-$&$-$&$-$&$-$& 12.6\\
V $+56\degr$60 &$149\fdg8$&$+5\fdg7$&B2.5\,III &$11.51^{*}$&$-$&$-$&$-$&$-$&13.7\\ 
V $+53\degr$22 &$151\fdg6$&$+2\fdg9$&B0\,Ia &9.32 &hi&       	 $3.21\pm0.13$  &  $0.90\pm0.03$  & $6.39\pm0.04$ &  13.0\\
\end{tabular}
\begin{list}{}{}
\item[]$^{(1)}$ The references for the $UBV$ photometry are 
(h) \citet{haug} or (hi) \citet{hil56}. The $JHK_{{\rm S}}$ photometry is from 2MASS. Two stars have no accurate $UBV$ photometry
and their distance moduli are estimates \citepalias[see][]{uno}. 
\end{list}
\end{center}
 \label{tab:data}
\end{table*}

The results of the fits with {\sc chorizos} 
%are listed in Table~\ref{tab:data}. They 
favour values of $R$ very close to the standard 
value $R=3.1$. There are few exceptions. LS~I~$+56\degr$92 is an
emission-line object \citep[cf.][]{neg04} and the high value of $R$
found reflects the presence of a  
circumstellar excess. The value adopted for the $DM$ of this object is an
approximation assuming standard reddening and no excess in $E(B-V)$
(formally, this should be taken as a lower limit, 
but is unlikely to differ much from the actual value).
The B5\,Ia supergiant \object{HD 25914} (LS~V~$+56\degr$56) also has a
very high value for $R$,  
suggesting a circumstellar excess as well. This is a known variable
(GQ~Cam), likely to be 
a very luminous star. Leaving aside these two objects, we have an
average $R=2.97\pm0.14$ (where the error represents the standard
deviation of the sample). This value is compatible within errors with
the standard value. Based on a photometric study of stars in the area,
\citet{zda05} concluded that $R=2.9$, in very good agreement with our
value, and that extinction is caused mostly by material at $<1.5$~kpc
from the Sun. This result justifies the assumption of standard
reddening used when dereddening the Str\"omgren photometry in
Section~\ref{sec:dist}. 
We note a tendency for double-lined stars to give lower values of $R$,
though there are too few such systems to know if this is a real trend.

With these
values, we calculate the distance modulus to each object. Since the
publication of \citetalias{uno}, several works have advocated for a
reduction in the temperature scale and absolute magnitude scale of
O-type stars. Because of this, we use the absolute magnitude scale of
\citet{martins} rather than that used in \citetalias{uno}. For B
spectral types, the calibration of \citet{hme84} has been used. In the
B0-B1 interval, the calibrations are not consistent, and an
interpolation has been made, resulting in the calibration given in
Table~\ref{tab:cal}. For the two luminosity class
Iab supergiants in the sample, we adopt  $M_{V}=-6.3$ for
LS~I~$+56\degr$99 (=\object{HD~237211}, O9.5\,Iab) and $M_{V}=-6.5$
for LS~I~$+55\degr$58 (= 
BD~$+55\degr$838, B2.5\,Iab). The values obtained are also listed in
Table~\ref{tab:data}.

\begin{table}[t]
\caption{Absolute magnitude calibration used here.}
  \begin{center}
\begin{tabular}{lcccccc}
& V & IV & III & II & Ib & Ia\\
\hline
O6 &$-4.9$&$-$ &$-5.7$&$-$& $-$& $-6.4$\\
O7 &$-4.6$&$-$ &$-5.5$&$-$& $-$& $-6.5$\\
O8 &$-4.3$&$-$ &$-5.4$&$-$& $-$& $-6.5$\\
O9 &$-4.1$&$-$ &$-5.3$&$-$& $-6.0$& $-6.5$\\
O9.5/7 &$-3.9$&$-$ & $-5.2$& $-$&$-6.0$&$-6.6$\\
B0 &$-3.8$&$-4.5$&$-5.1$&$-5.8$&$-6.0$&$-6.6$\\
B0.2 &$-3.8$&$-4.5$&$-5.1$&$-5.6$&$-6.0$&$-6.7$\\
B0.5 &$-3.5$&$-4.2$&$-5.0$&$-5.5$&$-6.0$&$-6.9$\\
B0.7 &$-3.5$&$-4.2$&$-4.8$&$-5.3$&$-6.0$&$-7.0$\\
B1 &$-3.2$&$-3.8$&$-4.3$&$-5.1$&$-6.0$&$-7.0$\\
B1.5 &$-2.8$&$-3.3$&$-3.9$&$-5.0$&$-5.8$&$-7.2$\\
B2 &$-2.5$&$-3.1$&$-3.7$&$-4.8$&$-5.8$&$-7.4$\\
B2.5 &$-2.0$&$-3.1$ &$-3.5$&$-4.8$&$-5.8$&$-7.4$\\
B3 &$-1.6$&$-2.5$&$-3.0$&$-4.7$&$-5.8$&$-7.2$\\
\end{tabular}
\end{center}
 \label{tab:cal}
\end{table}

With the values obtained, HD~237204 (LS~I~$+56\degr$98), is clearly
a foreground object, at a distance compatible with the Perseus Arm. 
The other 19 objects give $DM=12.9\pm0.5$ (1~$\sigma$). LS~I~$+55\degr$53 
has a distance modulus $>2\sigma$ shorter than the average and its spectral type (B3\,III) 
suggests a star significantly less massive than all the other ones. Therefore
we also take it for a foreground object. 
Leaving out this object, we find $DM=13.0\pm0.4$. 

\section{Discussion}

We have found the small cluster Alicante~1 around BD~$+56\degr$864 and
derived a distance modulus $DM=13.0\pm0.2$. This value is identical within the errors
to the average spectroscopic distance modulus to bright members of 
Cam~OB3. In view of this, we can firmly identify Alicante~1 as the first 
known cluster in Cam~OB3 and confirm the distance modulus to Cam~OB3 as 13.0, 
corresponding to 4.0~kpc.

\subsection{Alicante~1}
\label{sec:cluster}

As there are no evolved stars in Alicante~1, its age cannot be accurately
determined. However, the presence of mid O-type stars close to the
ZAMS forces it to be $<4$~Myr. On the other hand, the lack of any
\ion{H}{ii} region associated with the stars suggests that they must
have had time to disperse their maternal cloud entirely. Therefore an
age of 2\,--\,3~Myr is greatly favoured. At this age, all stars later
than $\sim$B8 ($\sim 3\,M_{\sun}$) must still be in the
pre-main-sequence phase.

The HR diagram of Alicante~1 (see Fig.~\ref{fig:real}) shows an
obvious deficiency of early B-type stars.  
Taking into account the binary nature of BD~$+56\degr$864 and the
very likely membership of BD~$+56\degr$866,  
there are at least four O-type stars in the cluster, three more
massive than $30\,M_{\sun}$. In contrast, 
the only early B-type stars are the two components of LS~I~$+57\degr$137. The
photometric sequence 
only starts at B3\,V and the number of B-type members is not very
large. This mass distribution is quite 
different from a standard IMF \citep{kroupa}. It may be argued that, 
in such small cluster, small number statistics may result in large
deviations from a standard value. However, 
it is tempting to assume that dynamical ejection from the
cluster core has played a role in shaping the observed IMF. Clusters
containing hard binaries with two components of similar mass may be
quite effective at ejecting stars via dynamical ejections \citep{ld90}, 
especially if massive stars are
mainly born as part of multiple systems \citep{pflam}. The
majority of stars ejected will be B-type stars and their ejection
velocity will be inversely proportional to their mass.

The presence of BD~$+56\degr$866 $\sim 5\arcmin$ to the South of the
two other O-type stars 
may represent evidence in favour of ejection, though the stellar
distribution in Alicante~1 may rather result from a spread of star
formation in small clumps. 
%While most stars in
%Cam~OB3 have measured (Tycho2) proper  
%motions compatible with zero (for instance, BD~$+56\degr$864 and
%LS~I~$+57\degr$137 have  
%P.M.($\alpha$, $\delta$) = ($-2.3\pm2.5$,$-0.3\pm2.4$) and
%($1.8\pm2.7$,$0.7\pm2.6$) mas, respectively), BD~$+56\degr$866 
%may have measurable motion, P.M.($\alpha$, $\delta$)
%=($5.1\pm3.3$,$6.8\pm3.2$) mas. 
The only other unevolved O-type star in Cam~OB3 is LS~I~$+57\degr$136
(O8.5\,V) which lies $\sim20\arcmin$ away from Alicante~1. At
$d=4.0$~kpc, this corresponds to a projected  
separation of $\approx23$~pc. With a typical runaway speed of
$10\:{\rm km}\,{\rm s}^{-1}$,  
a star would need 2.5~Myr to cover this distance and so a hypothetical ejection 
of LS~I~$+57\degr$136 from Alicante~1 is not
unreasonable. Unfortunately, there are no accurate 
measurements of proper motions for this star.

The region with a higher member density is moderately small, $\sim
2\arcmin$ across, which corresponds to $\approx2.3$~pc at
$d=4.0$~kpc. It contains 20 possible members earlier than $\sim$A2. It
is difficult to assess if the stars around the core represent a halo
or rather small groups that have formed in the vicinity of the small
cluster. This spatial distribution, with small concentrations of
early-type stars extended over a moderately large area, is not
uncommon in this area of the sky (cf. the discussion of clumps of
\ion{H}{ii} regions towards the Anticentre in \citealt{mof79}).

\subsection{Cam~OB3 and the Cygnus Arm}

The majority of the members of Cam~OB3 listed in Table~\ref{tab:data}
are evolved massive stars and hence older than Alicante~1. The cluster
is not the nuclear region of the association, but may simply represent the
latest small star-forming region to have actively created high-mass
stars. No other concentrations are obvious amongst members.

In their photometric study of this area, \citet{zda05} failed to find
any evidence of the Perseus Arm, either as an increase in extinction
or a concentration of stars. The lack of intervening material allows
the identification of Cygnus Arm tracers. Cam~OB3 is not a very
massive association, but with a radial extent $>100$~pc and a
significant number of massive stars, it is a reliable spiral
tracer. Nearby, \citet{mof79} estimate $DM=13.2\pm0.2$ for the young
open cluster \object{Waterloo 1}, at $l=151\fdg4$, in very good
agreement with the distance to Cam~OB3. \citet{mof79} also estimate
distances for two small groups of
young stars associated with the \ion{H}{ii} regions \object{Sh 2-217}
and \object{Sh 2-219} (at $l\approx159\fdg3$), both of which contain
embedded clusters \citep{deh03}, finding
$DM=13.1\pm0.3$ and $DM=13.6\pm0.3$, respectively. Therefore it seems
that the Cygnus Arm is well traced in this area (see discussion in
\citetalias{uno}). The young clusters observed in this area are not
very massive and there seems to be a high tendency to sparse groupings
of massive stars, rather than compact concentrations.

\section{Conclusions}

Our study of Cam~OB3 has resulted in 18 likely members from the
Luminous Star catalogue. When their reddenings are calculated
individually, they support a value of $R=3.0$, in good agreement with
the photometric determination of \citet{zda05}. Their average distance
modulus is $13.0\pm0.4$ in good agreement with previous determinations
based on a smaller number of stars.

Around the earliest members of the association, BD~$+56\degr$864
(O6\,V) and LS~I~$+57\degr$138 (O7\,V), we find a concentration
of B and early A stars, which clearly trace the main sequence of a
small and sparse cluster. The sequence also includes the nearby
LS~I~$+57\degr$137 (B1.5\,V) and BD~$+56\degr$866 (O9\,V) and some
other faint stars around them. We spectroscopically confirm the
brightest uncatalogued members of the sequence to be B3--5\,V
stars. We call this uncatalogued cluster Alicante~1 and find a
distance modulus of $13.0\pm0.2$ from ZAMS fitting.

These results definitely confirm the existence of Cam~OB3 as an
association in the Cygnus Arm and allow an accurate determination of
its distance as $4.0\pm0.4$~kpc.

\begin{acknowledgements}

We are very grateful to Jes\'us Ma\'{\i}z Apell\'aniz for advice on
the use and interpretation of the {\sc chorizos} code. We also thank
the referee for helpful comments that improved the paper.

This research has been partially supported by the Ministerio de Educaci\'on y
Ciencia under
grant AYA2005-00095 and by the Generalitat Valenciana under grant GV04B/729.  

  The INT is operated on the island of La
Palma by the Isaac Newton Group in the Spanish Observatorio del Roque
de Los Muchachos of the Instituto de Astrof\'{\i}sica de Canarias.
Based in part on observations made at Observatoire de Haute Provence 
(CNRS), France. The Nordic Optical Telescope is operated 
on the island of La Palma jointly by Denmark, Finland, Iceland,
Norway, and Sweden, in the Spanish Observatorio del Roque de los
Muchachos of the Instituto de Astrofisica de Canarias. The data were
taken with ALFOSC, which is 
owned by the Instituto de Astrof\'{\i}sica de Andaluc\'{\i}a (IAA) and
operated at the Nordic Optical Telescope under agreement
between IAA and the NBIfAFG of the Astronomical Observatory of
Copenhagen.

This research has made use of the Simbad data base, operated at CDS,
Strasbourg, France. This publication makes use of data products from
the Two Micron All 
Sky Survey, which is a joint project of the University of
Massachusetts and the Infrared Processing and Analysis
Center/California Institute of Technology, funded by the National
Aeronautics and Space Administration and the National Science
Foundation.     
      
\end{acknowledgements}

\clearpage
\begin{table}
\begin{minipage}[t]{\textwidth}
\caption{Coordinates ($XY$) in the map (Fig.~\ref{fig:full}) of likely
  members of Alicante~1 with $ubvy$ photometry. 
2MASS identification for these stars and their coordinates are also given.}
\label{tab:rawphot}
\centering
\begin{tabular}{cccccc}
\hline\hline
Number&RA (J2000)&DEC (J2000)& Name (2MASS)&$X$ (Pixels)&$Y$ (Pixels)\\
\hline
695$^{*}$&03 58 46.53    &+57 10 43.6  	&059.693873+57.178764&414.93	&1411.15\\
266&03 59 11.36 	&+57 14 10.5 	&059.797345+57.236237&975.86	&808.75\\
264&03 58 49.39 	&+57 14 17.2	&059.705783+57.238117&466.86	&801.75\\
570&03 59 17.33	&+57 08 52.9	&059.822194+57.148026&1135.31	&1708.00\\
439&03 58 44.97	&+57 11 29.6 	&059.687384+57.191555&376.07	&1280.98\\
280&03 59 03.00	&+57 14 04.2	&059.762506+57.234509&782.23	&831.13\\
609&03 59 15.96	&+57 08 18.4	&059.816485+57.138458&1106.08	&1806.73\\
271&03 59 40.68	&+57 14 52.1 	&059.919501+57.247795&1012.12	&814.95\\
223&03 59 14.29 	&+57 14 54.2	&059.809535+57.248375&1039.09	&682.36\\
298&03 59 19.05	&+57 13 41.0	&059.829361+57.228069&1154.34	&887.31\\
304&03 59 16.47	&+57 13 36.1 	&059.818637+57.226681&1095.23	&903.19\\
44&03 58 31.32	&+57 18 16.2	&059.630503+57.304493&34.41	&131.62\\
83&03 59 14.86	&+57 17 32.4 	&059.811910+57.292332&1040.82	&231.82\\
281&03 59 11.80	&+57 14 02.1 	&059.799157+57.233925&985.28	&831.91\\
245&03 59 14.89	&+57 14 33.8	&059.812057+57.242710&1054.73	&740.04\\
302&03 59 11.80 	&+57 13 39.7	&059.799169+57.227707&986.99	&895.67\\
278&03 59 12.72	&+57 14 04.1	&059.803008+57.234467&1006.64	&825.77\\
254&03 59 14.89	&+57 14 33.8	&059.812057+57.242710&936.50	&774.56\\
217&03 59 40.68	&+57 14 52.1 	&059.919501+57.247795&1648.13	&672.37\\
381&03 59 18.66	&+57 12 25.4	&059.827744+57.207047&1150.71	&1103.12\\
444&03 58 53.60	&+57 11 19.3 	&059.723338+57.188694&576.28	&1305.51\\
222&03 59 11.62 	&+57 14 54.8 	&059.798424+57.248566&977.58	&682.26\\
328&03 59 21.02 	&+57 13 15.9	&059.837604+57.221096&1210.75	&960.55\\
251&03 59 09.61	&+57 14 26.2	&059.790049+57.240608&933.08	&764.68\\
289&03 59 18.25	&+57 13 49.7	&059.826022+57.230476&1135.10	&863.31\\
649&03 59 14.10	&+57 07 37.4	&059.808753+57.127056&1065.84	&1924.29\\
297&03 59 15.23	&+57 13 42.2 	&059.813468+57.228401&1066.39	&886.79\\
321&03 59 05.77	&+57 13 22.5	&059.774047+57.222904&848.99	&948.38\\
351&03 58 49.63	&+57 12 49.2	&059.706797+57.213665&478.45	&1051.91\\
610&03 58 35.77 	&+57 08 23.5 	&059.649042+57.139870&175.46	&1815.13\\
62&03 58 45.35 	&+57 17 52.4	&059.688969+57.297894&359.40	&191.55\\
119&03 58 57.73	&+57 16 53.0	&059.740534+57.281399&648.78	&353.67\\
89&03 58 57.74	&+57 17 31.5 	&059.740576+57.292084&646.15	&244.04\\
635&03 58 34.70	&+57 08 01.3 	&059.644580+57.133690&152.05	&1879.19\\
230&03 58 54.97	&+57 14 51.8	&059.729049+57.247719&593.33	&700.30\\
396&03 58 34.71 	&+57 12 22.2	&059.644608+57.206169&135.28	&1136.67\\
694$^{*}$&03 59 07.49	&+57 14 11.7 	&059.781192+57.236591&883.50		&806.50\\
693$^{*}$&03 59 18.30	&+57 14 13.8	&059.826231+57.237160&1134.50		&795.50\\
\hline
\end{tabular}
\begin{list}{}{}
\item[]$^{(*)}$ Catalogued stars: \#693 = BD$+56\degr$864, \#694 =
  LS~I~$+57\degr$138, \#695 =  LSI~$+57\degr$137
\end{list}
\end{minipage}
\end{table}

\clearpage
\begin{table}
\caption{ Photometry of the stars with numbers in Fig.~\ref{fig:full}}
\label{table:4}
\centering
\begin{tabular}{cccccccc}
\hline\hline
 Number&V&(b-y)&$c_{1}$&$c_{0}$&$M_{V}$&$E(b-y)$&$E(c_{1})$\\
\hline
695	&11.171	&0.325	&0.176&	0.088	&	$-3.723$&	0.441	&	0.088\\	
266	&	13.777	&	0.377	&	0.441&	0.347	&	$-1.235$&	0.468	&	0.094\\	
264	&	14.066	&	0.423	&	0.536&	0.435	&	$-1.110$&	0.506	&	0.101\\	
570	&	14.094	&	0.401	&	0.461&	0.363	&	$-1.016$&	0.491	&	0.098\\
439	&	14.295	&	0.398	&	0.528&	0.432	&	$-0.773$&	0.481	&	0.096\\
280	&	14.394	&	0.365	&	0.548&	0.459	&	$-0.518$&	0.445	&	0.089\\
609	&	14.635	&	0.408	&	0.747&	0.653	&	$-0.383$&	0.469	&	0.094\\
271     &       14.684  &       0.437   &       0.578&  0.475	&	$-0.537$&	0.516	&	0.103\\
223	&	14.689	&	0.397	&	0.683&	0.590	&	$-0.307$&	0.464	&	0.093\\
298	&	14.796	&	0.374	&	0.584&	0.494	&	$-0.140$&	0.450	&	0.090\\
304	&	15.029	&	0.404	&	0.773&	0.681	&	0.040	&	0.462	&	0.092\\
44	&	15.232	&	0.390	&	0.563&	0.469	&	0.215	&	0.469	&	0.094\\
83      &       15.284  &       0.456   &       1.130&  1.034	&	0.218	&	0.481	&	0.096\\
281	&	15.339	&	0.428	&	1.132&	1.042	&	0.399	&	0.451	&	0.090\\
245	&	15.399	&	0.418	&	0.952&	0.860	&	0.425	&	0.459	&	0.092\\
302	&	15.454	&	0.420	&	1.092&	1.003	&	0.532	&	0.447	&	0.089\\
278	&	15.514	&	0.464	&	1.125&	1.027	&	0.410	&	0.489	&	0.098\\
254	&	15.523	&	0.380	&	0.920&	0.835	&	0.705	&	0.423	&	0.085\\
217	&	15.549	&	0.488	&	1.150&	1.048	&	0.349	&	0.512	&	0.102\\
381	&	15.585	&	0.441	&	0.961&	0.865	&	0.513	&	0.482	&	0.096\\
444	&	15.590	&	0.484	&	0.901&	0.794	&	0.299	&	0.533	&	0.107\\
222	&	15.674	&	0.448	&	0.936&	0.838	&	0.559	&	0.492	&	0.098\\
328	&	15.753	&	0.431	&	1.037&	0.944	&	0.758	&	0.464	&	0.093\\
251	&	15.853	&	0.461	&	1.117&	1.020	&	0.759	&	0.487	&	0.097\\
289	&	15.854	&	0.440	&	1.136&	1.043	&	0.862	&	0.463	&	0.093\\
649	&	15.891	&	0.496	&	0.941&	0.833	&	0.564	&	0.541	&	0.108\\
297	&	15.892	&	0.433	&	1.035&	0.942	&	0.887	&	0.466	&	0.093\\
321	&	15.943	&	0.444	&	1.024&	0.928	&	0.885	&	0.479	&	0.096\\
351	&	15.991	&	0.463	&	1.049&	0.950	&	0.858	&	0.496	&	0.099\\
610	&	16.005	&	0.479	&	1.143&	1.042	&	0.842	&	0.503	&	0.101\\
62      &	16.253	&	0.483	&	0.851&	0.744	&	0.945	&	0.537	&	0.107\\
119	&	16.343	&	0.467	&	1.040&	0.940	&	1.189	&	0.501	&	0.100\\
89	&	16.732	&	0.501	&	1.088&	0.982	&	1.447	&	0.532	&	0.106\\
635	&	16.893	&	0.461	&	0.992&	0.892	&	1.745	&	0.500	&	0.100\\
230     &       16.966  &       0.479   &       1.088&  0.986   &       1.779   &       0.509   &       0.102\\
396	&	17.026	&	0.474	&	1.098&	0.998	&	1.866	&	0.502	&	0.100\\
694	&	10.006	&	0.305	&	0.054&	$-0.038$&	$-5.012$&	0.450	&	0.115\\
693	&	9.600	&	0.311	&	0.040&	$-0.052$&	$-5.418$&	0.450	&	0.100\\

\hline
\end{tabular}
\end{table}

\end{document}